\documentstyle[preprint,aps,tighten]{revtex}

\begin{document}
\title{Tetraquarks in a chiral constituent quark model}
\author{J. Vijande$^{1}$, F. Fern\'{a}ndez$^{1}$, 
A. Valcarce$^{1}$, B. Silvestre-Brac $^{2}$}
\author{$^{1}$ Grupo de F\'{\i}sica Nuclear, 
Universidad de Salamanca, E-37008 Salamanca, Spain}
\author{$^{2}$ Institut des Sciences Nucl\'eaires,
53 Avenue des Martyrs, F-38026 Grenoble Cedex, France}
\maketitle

\begin{abstract}
We analyze the possibility of heavy-light 
tetraquark bound states by means of 
a chiral constituent quark model.
The study is done in a variational approach.
Special attention is paid to the contribution given by
the different terms of the interacting potential and also
to the role played by the different color channels.
We find a stable state for both $qq\bar{c}\bar{c}$ and $qq\bar{b}\bar{b}$
configurations. Possible decay modes of these structures are analyzed.
\end{abstract}

\vspace*{2cm}
\noindent Pacs: 12.39.Jh, 12.39.Pn

\newpage

\section{Introduction}

The potentiality of the quark model for 
hadron physics in the low-energy regime
was manifest when it was used to classify
all the known hadron states. Describing
hadrons as $q\overline{q}$ or $qqq$ configurations, 
their quantum numbers were correctly explained.
This assignment was based on the comment by Gell-Mann \cite{gell}
introducing the notion of quark: 
{\it 'It is assuming that the lowest baryon
configuration ($qqq$) gives just the 
representations 1, 8 and 10, that
have been observed, while the lowest meson 
configuration ($q \bar{q}$)
similarly gives just 1 and 8'}. Since then, it is assumed 
that these are the two 
configurations involved in the description
of physical hadrons. However, color confinement 
is also compatible with other multiquark 
structures like the tetraquark $qq \bar{q}\bar{q}$
introduced by Jaffe \cite{jaf}. 
During the last two decades of the past century
there appear a number of experimental data that 
are hardly accommodated in the traditional scheme defined
by Gell-Mann. Besides, new experiments are being designed
looking for these unusual structures. 

The possible existence of tetraquarks has been suggested 
in two different scenarios. 
The first one is a system composed of two light quarks (antiquarks)
and two heavy antiquarks (quarks). 
Although such an object is experimentally difficult to produce and also to detect 
\cite{new} it has been argued that for sufficiently large 
heavy quark mass the tetraquark system should be bound \cite{mar}. 
The second scenario corresponds to the 
scalar mesons, $J^{PC}=0^{++}$, where a huge amount of
experimental data is available nowadays. However, in spite of that the
situation is not yet conclusive.  From the theoretical point of
view the scalar sector is very particular, 
because to obtain a positive parity state from a $q\overline{q}$ 
pair one needs at least one unit of
angular momentum. Apparently this costs an energy around 1 GeV, since
similar meson states ($1^{++}$ and $2^{++}$) lie 
above 1.2 GeV. However, a $q^{2}\bar{q}^{2}$ state can couple 
to $J^{P C}=0^{++}$ without orbital
excitation and, as a consequence, it could be a 
serious candidate to describe the
lightest scalar mesons. In this paper we focus our attention on the first subject,
the heavy-light tetraquarks, whose existence has been recently examined
\cite{boris} from experimental data of the SELEX collaboration \cite{selex}.

The stability of the $qq \bar{Q}\bar{Q}$ system relies on the 
mass of the heavy quark \cite{mar}. The heavier the quark the stronger 
the short-range Coulomb attraction, in such a way that it could
play a decisive role to bind
the tetraquark system. The binding energy of a $1/r$ potential is
proportional to the mass and it 
must lead to a bound state in the limit of
infinite quark mass. Moreover the $\bar Q \bar Q$ pair 
brings a small kinetic energy into the system and then contributes 
to stabilize it. On the other hand, heavy-light tetraquarks are
ideal systems to study the interplay between the different quark
interactions because chiral symmetry is spontaneously broken 
in the light sector but it is explicitly broken  in
the heavy one. 

Heavy-light tetraquarks have been studied in the past in
different ways. Carlson {\it et al.} \cite{hell} used a 
potential derived from the
MIT bag model in the Born-Oppenheimer approximation. 
The calculations were done by means of the Green's
function Monte Carlo Method. They found a $J^P=1^+$ 
isoscalar $bb\bar{u} \bar{d}$ bound state and 
they also concluded that the $cc\bar{u}\bar{d}$ is not bound.
A different approach was followed by Manohar and Wise 
\cite{wise} who studied systems of 
two heavy-light ($Q \bar q$) mesons interacting by a
potential determined at long distances 
by a one-pion exchange computed using
Chiral Perturbation Theory. 
They found that for $Q=b$ this long range
potential may be sufficiently attractive to 
produce a weakly bound two-meson
state, although states where $Q=c$ are not bound. 
In the framework of the nonrelativistic
quark potential models, Silvestre-Brac and Semay
\cite{sils} have studied possible tetraquark 
structures using different parametrizations
of the Bhaduri potential \cite{bhad}. 
They found several bound state
candidates in the $bb\bar{q}\bar{q}$ 
and $bc\bar{q}\bar{q}$
configurations but not for the $cc\bar{q}\bar{q}$
structure. A different conclusion is obtained 
by Pepin {\it et al.} \cite{rich} using
a pseudoscalar meson-exchange interaction 
coming from the breaking of chiral symmetry
instead of the chromomagnetic potential.
Their results indicate that such interaction binds 
the heavy tetraquark systems
both for $Q=b$ and $Q=c$. 
Therefore, the theoretical situation seems to be
uncertain depending on whether chromomagnetic
or chiral interactions are used. Even in the case 
of the same kind of
interaction (one-pion exchange) the results of 
Refs. \cite{wise} and \cite{rich} are different.

Such a model dependence of the possible existence
of tetraquarks claims for calculations with interactions constrained in other
sectors. For this purpose, the chiral constituent
quark model of Ref. \cite{paco} is an ideally suited starting point. 
It is based on the idea that between the 
chiral symmetry breaking scale and the confinement scale,
QCD may be formulated for the light quark
sector as an effective theory of constituent 
quarks interacting through gluons and Goldstone 
modes associated to the spontaneous breaking of $SU(2)$ chiral symmetry.
Its parameters have been determined in the description of 
non-strange two- and three-baryon systems and
the hadron spectra \cite{paco,bruno,luis}. For the present study one needs to
include strange and heavy flavors and therefore the model has to be generalized, fixing the new
parameters in the meson spectra. This interaction will be used to solve the
Schr\"{o}dinger equation for the tetraquark system using a variational
method. For the spatial wave function we assume a linear
combination of gaussians and we will consider 
the two possible color configurations:
$\{\overline{3}3\}$ and $\{6\overline{6}\}$.

The paper is organized as follows. 
In Sec. II we describe the chiral constituent quark model. 
The method and details of the calculation are shown in Sec. III. Section IV 
is devoted to the analysis and discussion of the results.
Finally, a summary is presented in Sec. V.

\section{The Chiral Constituent Quark Model}

Since the origin of the quark model
hadrons have been considered to be built
by constituent (massive) quarks. Nowadays it is widely
recognized that the constituent quark mass, 
very different from the current quark
mass of the QCD lagrangian, appears 
because of the spontaneous breaking of the original 
$SU(3)_{L} \otimes SU(3)_{R}$ chiral symmetry 
at some momentum scale. The picture of
the QCD vacuum as a dilute medium of 
instantons \cite{diak} explains nicely
such a symmetry breaking, which is the most 
important nonperturbative phenomenon for
hadron structure at low energies. 
Quarks interact with fermionic zero modes
of the individual instantons in the medium and therefore
the propagator of a light
quark gets modified and quarks acquire a 
momentum dependent mass which
drops to zero for momenta higher than the inverse 
of the average instanton
size $\overline{\rho }$. The momentum 
dependent quark mass acts as a natural
cutoff of the theory. In the domain of 
momenta $k<1/\overline{\rho }$, a
simple lagrangian invariant under the 
chiral transformation can be derived as 
\cite{diak}

\begin{equation}
L=\overline{\psi }(i \gamma^\mu \partial_\mu -MU^{\gamma _{5}})\psi
\end{equation}

\noindent where $U^{\gamma _{5}}=\exp (i\pi ^{a}\lambda ^{a}\gamma _{5}/f_{\pi })$. 
$\pi ^{a}$ denotes the pseudoscalar fields 
$(\vec{\pi } ,K_{i},\eta_8)$ with i=1,...,4, and $M$ is the constituent quark mass. An 
expression of the constituent quark mass can be obtained
from the theory, but it also can be parametrized as $M(q^{2})=m_{q}F(q^{2})$ with

\begin{equation}
F(q^{2})=\left[ \frac{\Lambda^{2}}{\Lambda^{2}+q^{2}}
\right] ^{\frac{1}{2}}
\end{equation}

\noindent
where $\Lambda$ determines the scale 
at which chiral symmetry is broken.
Even if one does not believe in instantons 
as the microscopic mechanism of
spontaneous chiral symmetry breaking,
one has to admit that once a constituent
quark mass is generated by some mechanism 
such quarks inevitably have to
interact through Goldstone modes. Whereas the lagrangian 
$\overline{\psi } (i\gamma^\mu \partial_\mu -M)\psi $ 
is not invariant under chiral rotations, the
lagrangian of Eq. (1) is invariant since the rotation 
of the quark fields can be
compensated renaming the bosons fields.
$U^{\gamma _{5}}$ can be expanded in terms of boson fields as,

\begin{equation}
U^{\gamma _{5}}=1+\frac{i}{f_{\pi }}\gamma ^{5}\lambda ^{a}\pi ^{a}-\frac{1}{%
2f_{\pi }^{2}}\pi ^{a}\pi ^{a}+...
\end{equation}
The first term generates the quark constituent mass and the
second one gives rise to a one-boson 
exchange interaction between quarks. The
main contribution of the third term comes 
from the two-pion exchange which
will be simulated by means of the one-sigma exchange potential.
Based on the non-relativistic approximation of the above lagrangian, and making
use of the physical $\eta$ instead the octect one (this is the reason why a
mixing angle $\theta_p$ appears), one can write the following potentials
between quarks,

\begin{equation}
V_{PS}(\vec{r}_{ij})={\frac{g_{ch}^{2}}{{4\pi }}}{\frac{m_{PS}^{2}}{{
12m_{i}m_{j}}}}{\frac{\Lambda _{PS}^{2}}
{{\Lambda _{PS}^{2}-m_{PS}^{2}}}}
m_{PS}\left[ Y(m_{PS}\,r_{ij})-
{\frac{\Lambda _{PS}^{3}}{m_{PS}^{3}}}
Y(\Lambda _{PS}\,r_{ij})\right] (\vec{\sigma}_{i}\cdot \vec{\sigma}
_{j})\sum_{a=1}^{3}{(\lambda_{i}^{a}\cdot \lambda_{j}^{a})} \, ,
\label{ps}
\end{equation}
\begin{equation}
V_{S}(\vec{r}_{ij})=-{\frac{g_{ch}^{2}}{{4\pi }}}
{\frac{\Lambda _{S}^{2}}{{
\Lambda _{S}^{2}-m_{S}^{2}}}}m_{S}
\left[ Y(m_{S}\,r_{ij})-{\frac{\Lambda _{S}
}{m_{S}}}Y(\Lambda _{S}\,r_{ij})\right] ,
\end{equation}
\begin{equation}
V_{K}(\vec{r}_{ij})={\frac{g_{ch}^{2}}{{4\pi }}}{\frac{m_{K}^{2}}{{
12m_{i}m_{j}}}}{\frac{\Lambda _{K}^{2}}
{{\Lambda _{K}^{2}-m_{K}^{2}}}}m_{K}
\left[ Y(m_{K}\,r_{ij})-{\frac{\Lambda _{K}^{3}}{m_{K}^{3}}}Y(\Lambda
_{K}\,r_{ij})\right] (\vec{\sigma}_{i}\cdot 
\vec{\sigma}_{j})\sum_{a=4}^{7}{(
\lambda_{i}^{a}\cdot \lambda_{j}^{a})} \, ,
\end{equation}
\begin{equation}
V_{\eta }(\vec{r}_{ij})={\frac{g_{ch}^{2}}{{4\pi }}}
{\frac{m_{\eta }^{2}}{{
12m_{i}m_{j}}}}{\frac{\Lambda _{\eta}^{2}}{{\Lambda _{\eta}^{2}-m_{\eta
}^{2}}}}m_{\eta }\left[ Y(m_{\eta }\,r_{ij})-
{\frac{\Lambda _{\eta}^{3}}{
m_{\eta }^{3}}}Y(\Lambda _{\eta}\,r_{ij})\right] (\vec{\sigma}_{i}\cdot 
\vec{\sigma}_{j})\left[ cos\theta _{p}
(\lambda_{i}^{8}\cdot \lambda
_{j}^{8})-sin\theta _{p} \right]
\end{equation}

\noindent
where $g_{ch}=m_q/f_{\pi}$, the $\lambda's$ are the $SU(3)$ flavor 
Gell-Mann matrices, and $Y(x)$ is the standard Yukawa function.
The chiral coupling constant $g_{ch}$ is related to the $\pi NN$
coupling constant by 
\begin{equation}
\frac{g_{ch}^2}{4\pi }=\left( \frac{3}{5}
\right)^{2}{\frac{g_{\pi NN}^{2}}{{
4\pi }}}{\frac{m^2_{u,d}}{m_{N}^{2}}}
\end{equation}
Proceeding on this way one assumes that flavor $SU(3)$
is an exact symmetry, only broken by the different 
mass of the strange quark.
$m_i$ is the quark mass, 
$m_{PS}$, $m_K$ and $m_\eta$ are the masses of the 
$SU(3)$ Goldstone bosons, taken to be their experimental values, and $m_S$ is
taken from the PCAC relation $m_S^2\sim m_{PS}^2+4\,m_{u,d}^2$ \cite{scad}.
In the heavy quark sector, chiral symmetry is explicitly
broken and therefore these interactions will not appear.

For higher momentum transfer quarks still interact through gluon
exchanges. De R\'{u}jula {\it et al.} \cite{DeR} 
proposed that gluon exchange between constituent quarks can be described as an
effective interaction according to the lagrangian

\begin{equation}
L_{gqq}=i\sqrt{4\pi }\alpha _{s}\overline{\psi }\gamma _{\mu }G^{\mu
}\lambda^{c}\psi
\end{equation}
where $\lambda^{c}$ are the $SU(3)$ color matrices
and $G^{\mu }$ is the gluon field. Using a nonrelativistic 
reduction one obtains coulomb and contact potentials that can be
parametrized as follows \cite{isg2}:

\begin{equation}
V_{OGE}(\vec{r}_{ij})={\frac{1}{4}}
\alpha _{s}\,\vec{\lambda^c}_{i}\cdot \vec{
\lambda^c}_{j}\,\left\{ {\frac{1}{r_{ij}}}-
{\frac{2\,\pi}{3m_{i}m_{j}}}
\vec{\sigma}_{i}\cdot \vec{\sigma}_{j}
\delta (\vec{r}_{ij})\right\}.
\label{oge}
\end{equation}

In order to obtain a unified description of light, strange and heavy mesons, a running
strong coupling constant has to be used \cite{isg2}. The standard expression for
$\alpha_s(Q^2)$ diverges when $Q\rightarrow\Lambda_{QCD}$ and therefore the coupling
constant has to be frozen at low energies \cite{sim}. We parametrize this behavior by
means of an effective scale dependent strong coupling constant similar to the one used in Refs.
\cite{pre2,coup}

\begin{equation}
\alpha_s(\mu)={\alpha_0\over{ln\left({{\mu^2+\mu^2_0}\over\Lambda_0^2}\right)}},
\label{asf}
\end{equation}

\noindent where $\mu$ is the reduced mass of the $q\bar q$ system and $\alpha_0$, $\mu_0$ and $\Lambda_0$ 
are fitted parameters. This equation gives rise to $\alpha_s\sim0.54$ for the light quark sector, 
a value consistent with the one used in the study of the nonstrange hadron phenomenology \cite{paco,bruno,luis}, and
it also has an appropriate high $Q^2$ behavior, $\alpha_s\sim0.127$ at the $Z_0$ mass \cite{pre1}. In order 
to avoid an unbound spectrum from below the delta function has to be regularized. Taken 
into account that for a coulombic system the typical size scales with the reduced
mass, we use a flavor-dependent regularization $r_0(\mu)={{\hat r_0}/{\mu}}$,

\begin{equation}
\delta({\vec r}_{ij}) \, \Rightarrow \,
{1 \over {4 \pi r_0^2(\mu)}} \,\,
{{e^{-r_{ij}/r_0(\mu)}} \over r_{ij}}.
\end{equation}

The other nonperturbative property of QCD, 
which cannot be explained by the
instanton liquid model, is confinement. 
Up to now it still remains a problem
to derive this property from QCD in an analytic 
manner. The only indication
we have on the nature of confinement is 
through lattice QCD studies. These
calculations show that $q\overline{q}$ 
systems are well reproduced at short
distances by a linear potential. This potential can be physically
interpreted in a picture in which the quark 
and the antiquark are linked
with a one-dimensional color flux tube 
or string with a string tension 
$\sigma _{q\overline{q}}$ and hence the $q\overline{q}$ potential is
proportional to the distance between 
the quark and the antiquark. However, pair creation screens the potential at
large distances \cite{scre}. A screened potential simulating the results of
lattice calculations is given by

\begin{equation}
V_{CON}(\vec{r}_{ij})=\{-a_{c}\,(1-e^{-\mu_c\,r_{ij}})+
\Delta\}(\vec{\lambda^c}_{i}\cdot \vec{ \lambda^c}_{j})\,
\end{equation}

\noindent
where $\Delta$ is a global constant to fit the origin of energies. At short
distances this potential presents a linear behavior with an effective
confinement strength $a=a_c\,\mu_c\,(\vec{\lambda^c}_{i}\cdot \vec{\lambda^c}_{j})$ 
while it becomes constant at large distances.

Let us resume the different pieces of the interacting potential
as a function of the quarks involved: 

\begin{equation}
V_{ij}= \left\{ 
\begin{array}{ll}
(ij)=(qq) \Rightarrow V_{CON}+V_{OGE}+V_{PS}+V_{S}+V_{\eta } &  \\ 
(ij)=(qs) \Rightarrow V_{CON}+V_{OGE}+V_{S}+V_{K}+V_{\eta } &  \\ 
(ij)=(ss) \Rightarrow V_{CON}+V_{OGE}+V_{S}+V_{\eta } &  \\ 
(ij)=(qQ) \Rightarrow V_{CON}+V_{OGE} &  \\ 
(ij)=(QQ) \Rightarrow V_{CON}+V_{OGE} & 
\end{array}
\right.
\label{pot}
\end{equation}

\noindent
The corresponding $q \bar{q}$ potential is obtained from the 
$qq$ one as detailed in Ref. \cite{bumu}. In the case of $V_K(\vec{r}_{ij})$,
where G-parity is not well defined, the transformation 
is given by $\lambda^a_1 \cdot \lambda^a_2 \to
\lambda^a_1 \cdot (\lambda^a_2)^T$, which recovers the standard
change of sign in the case of the pseudoscalar exchange between
two nonstrange quarks.
Assuming a nonrelativistic expression for the kinetic energy,
the hamiltonian of the system takes the form:

\begin{equation}
H=\sum\limits_{i}\left( m_{i}+\frac{\vec p_{i}^{\,2}}{2m_{i}}\right)
+\sum_{i<j}V_{ij}
\label{ham}
\end{equation}
where $V_{ij}$ is given by Eq. (\ref{pot}).

\section{Tetraquark wave function}

For the description of the $qq\bar{Q}\bar{Q}$ 
system we introduce the Jacobi coordinates

\begin{eqnarray}
\text{ \ \ \ \ \ \ \ \ \ \ }\vec{x} &=&\vec{r}_{1}-
\vec{r}_{2}\text{ \ \ \ \ \ \ \ \ \ \ \ \ \ \ \ }\vec{y
}=\vec{r}_{3}-\vec{r}_{4} \\
\vec{z} &=&\frac{m_{1}\vec{r}_{1}+m_{2}\vec{
r}_{2}}{m_{1}+m_{2}}-\frac{m_{3}\vec{r}_{3}+m_{4}\vec{r
}_{4}}{m_{3}+m_{4}}\text{ \ \ \ \ \ \ \ \ \ \ \ \ \ }\vec{R}=
\frac{\sum m_{i}\vec{r}_{i}}{\sum m_{i}}
\end{eqnarray}
where particles $1$ and $2$ are quarks and $3$ and $4$ antiquarks.
The four-body problem is solved using a 
wave function that includes all 
the possible flavor-spin-color channels that
contribute to a given configuration. For each channel $s$, such
a wave function will be a tensor product of
a color ($g_s$), flavor ($f_s$), spin ($\chi_s$) and spatial 
($R_s$) wave functions

\begin{equation}
\mid \phi _{s}>=g_{s} (1234) f_s (1234) \chi _{s} (1234)
R_{s} (1234)
\end{equation}

Concerning the spatial wave function, the
most general one with $L=0$ may 
depend on six scalar quantities

\begin{equation}
R_s(1234)=R_s(x^{2},y^{2},z^{2},\vec{x}\cdot 
\vec{y,}\vec{x}\cdot \vec{z,}
\vec{y}\cdot \vec{z})
\end{equation}
Most part of tetraquarks studies have been done
by means of a completely symmetric spatial
wave function, depending only on 
the quadratic terms, $x^{2},y^{2} $ and $z^{2}$. 
This is a restricted choice which 
we will analyze in our calculation.
We define our variational 
spatial wave function as a linear combination of gaussians

\begin{equation}
R_s(1234)=
\sum_{i=1}^{n} \beta_{s}^{(i)} R_s^{(i)}=
\sum_{i=1}^{n} \beta_{s}^{(i)}
e^{-a^{(i)}_s \vec x^{\,2}-b^{(i)}_s \vec y^{\,2}-c^{(i)}_s \vec 
z^{\,2}-d^{(i)}_s \vec x \vec y -e^{(i)}_s \vec 
x \vec z -f^{(i)}_s \vec y \vec z}
\label{wave}
\end{equation}
where $n$ is the number of gaussians we use to expand the spatial
wave function of each color-spin-flavor component.

With respect to the color wave function,
one can couple the two quarks 
$(1,2)$ and the two antiquarks $(3,4)$ 
to color singlet in different ways

\begin{eqnarray}
&\mid &1_{13},1_{24}> \,\,\, , \,\,\, \mid 8_{13},8_{24}> \\
&\mid &1_{14},1_{23}> \,\,\, , \,\,\, \mid 8_{14},8_{23}> \\
&\mid &\overline{3}_{12},3_{34}> \,\,\, , 
\,\,\, \mid 6_{12},\overline{6}_{34}>
\end{eqnarray}
The first two couplings are convenient for asymptotic meson-meson
channels (or meson-meson molecules) 
while the third one is more appropriate
for tetraquark bound states. With our choice of the Jacobi coordinates
the last color basis results to be more suitable and essentially to treat the
Pauli principle in an easy way.

The spin part of the wave function
can be written as

\begin{equation}
\mid S_i>=\left[ (12)_{S_{12}}(34)_{S_{34}}\right] _{S}
\end{equation}
where the spin of the two quarks is coupled to $S_{12}$ and that
of the antiquarks to $S_{34}$.
Then we will have the following basis vectors as a function
of the total spin S=0,1,2:

\begin{equation}
S=0 \Rightarrow
\left\{ 
\begin{array}{ll}
\mid S_{1}>=\left[ (12)_{0}(34)_{0}\right] _{0} & \\
\mid S_{2}>=\left[ (12)_{1}(34)_{1}\right] _{0} &
\end{array}
\right.
\end{equation}
\begin{equation}
S=1 \Rightarrow
\left\{ 
\begin{array}{ll}
\mid S_{3}>=\left[ (12)_{0}(34)_{1}\right] _{1} & \\
\mid S_{4}>=\left[ (12)_{1}(34)_{0}\right] _{1} & \\
\mid S_{5}>=\left[ (12)_{1}(34)_{1}\right] _{1} &
\end{array}
\right.
\end{equation}
\begin{equation}
S=2 \Rightarrow
\left\{ 
\begin{array}{ll}
\mid S_{6}>=\left[ (12)_{1}(34)_{1}\right] _{2} &
\end{array}
\right.
\end{equation}

Concerning the flavor part, we consider light quarks
those with flavor $u$ and $d$ and heavy ones 
those with flavor $s$,
$c$ and $b$. The heavy quarks have
isospin zero so they do not contribute to the total isospin. 
Therefore we can classify the tetraquark wave function by 
the isospin of the light quarks $I=0,1$.
Taken into account all degrees of freedom, the Pauli principle must be
satisfied for each subsystem of identical quarks (antiquarks). 
It imposes restrictions on the quantum numbers of the basis 
states, which is the justification to use the coupling
$[(qq)(\bar Q \bar Q)]$.

Using the wave function described above, 
we search for a variational solution
of the hamiltonian of Eq. (\ref{ham}). 
The spin, color and flavor parts are integrated out
and the $\beta_s^{(i)}$ coefficients of the spatial wave function are
obtained by solving the system of linear equations 

\begin{equation}
\sum_s \sum_{i=1}^n \beta_s^{(i)} 
\, [\langle R_{s'}^{(j)}|\,H\,|R_s^{(i)}
\rangle - E\,\langle
R_{s'}^{(j)}|R_s^{(i)}\rangle \delta_{s,s'} ] = 0 
\qquad \qquad \forall \, j,s'
\end{equation}

\noindent 
once the eigenvalues  $E(a^{(i)}_s, ... ,f^{(i)}_s)$ are obtained
using a minimization procedure. In practice, we use
the MINUIT package \cite{minu} to obtain the energies.
The stable tetraquark states are identified by comparing the obtained
eigenvalues with the corresponding meson-meson threshold calculated 
with the same hamiltonian.

\section{Results}

As we have already discussed, the parameters appearing in our hamiltonian 
corresponding
to the light sector are fixed from the $NN$ interaction \cite{paco}. However 
there are others which cannot be determined in this way either because the 
$NN$ interaction does not depend on them (e.g. the confinement parameters) 
or because they appear as a consequence of the
generalization to strange and heavy flavors. Among the thoughtful 
criteria for fixing these remaining parameters the most commonly
used one is to fit the meson spectra. Being the meson masses the ones that
will govern the tetraquark thresholds and therefore will determine
if the tetraquark is bound or not, we think that a good fit of meson spectra
must be the most important criterium. 
This criterium is the one proposed long ago by Isgur and
collaborators \cite{isg} and also by Manohar and Wise \cite{wise}, and it has been
also adopted in many other works, which use the Bhaduri
potential whose parameters are fitted to charmonium \cite{sils}. Therefore, we have performed fits of
the $q\bar q$ sector using the tetraquark wave functions in the limit where
the two component mesons are isolated, what is equivalent to solve the $q\bar q$ system \cite{isg}.
The complete set of parameters is shown in
Table \ref{tabI} and the corresponding meson ground state masses are given in Table \ref{tabII}.

Tetraquarks will be stable under the 
strong interaction if their total energy
lies below all the possible, and allowed, 
two-meson thresholds. Therefore we will use the quantity

\begin{equation}
\Delta E=E_{T}(qq\bar{Q}\bar{Q})-E_{m_{1}}(q\bar{Q}
)-E_{m_{2}}(q\bar{Q})
\label{binding}
\end{equation}
to discriminate the stable tetraquark states.

We focus our attention on the lowest $L=0$, positive parity, states of the $qq\bar{Q}\bar{Q}$ 
system with $Q=b,c,s$ and $q=u,d$ for all possible
spin-isospin combinations. As a first step we consider a 
spatial wave function with a single gaussian 
and we compare the results obtained for the tetraquark energies
using the most general wave function [see Eq. (\ref{wave})]
or a completely symmetric spatial wave function. The energy
difference obtained are always smaller than 1\%.  Therefore we will use for 
the spatial wave function combinations of the quadratic
terms, $R(x^{2},y^{2},z^{2})$, except for those
spin-isospin channels which are Pauli forbidden 
if we use such a symmetric spatial wave function. 
For these channels, which correspond to $(S, I)=
(0,0)$ and $(2,0)$, we make the calculation with the full wave
function.

The variational calculation is done using as spatial wave function a sum of
gaussians. We start with one gaussian and we increase its number until
convergence in the tetraquark energy is reached. The typical number of gaussians needed
is six. 

The results are presented in Table \ref{tabIII}. Our calculation predicts two bound states, one
for the $qq\bar b \bar b$ system and one for the $qq\bar c \bar c$, in both
cases with $(S,I)=(1,0)$. No bound states are found for the
$qq\bar s \bar s$ system, the $(S,I)=(1,0)$ channel being again the most attractive one.
The other $(S,I)$ channels are strongly repulsive, specially those where the
Pauli principle forbids the spatially symmetric wave function, except for the $(S,I)=(1,1)$ 
and $(2,1)$ for the $qq\bar b\bar b$ system.
Comparing with the existing literature we find two
different types of calculations: those including only confinement and one-gluon
exchange, Refs. \cite{sils} and \cite{brink} and those
considering different models of Goldstone boson exchanges between quarks, 
Refs. \cite{wise} and \cite{rich}. Models including only
one-gluon exchange predict systematically a bound state in the $(S,I)=(1,0)$
channel for the $qq\bar b \bar b$ system and no bound state for the $qq\bar c\bar
c$ system. Among the others one finds different conclusions. In Ref.
\cite{wise} a weakly bound state is found only in the $qq\bar b\bar b$ system,
whereas Ref. \cite{rich} reports bound states for both $qq\bar b\bar b$ and
$qq\bar c\bar c$.

To clarify the importance of the different pieces of the interacting
hamiltonian we have redone our calculations switching-off the pseudoscalar part, Eq.(\ref{ps}), of the 
Goldstone boson exchanges. The results are shown in Table \ref{tabIV}. A first
glance to the results tells us that the bound states have disappeared. This
part of the interaction plays a relevant role to bind the
$(S,I)=(1,0)$ channel \cite{rich} and it also favors the binding in the $(S,I)=(0,0)$ but not
in the other channels. One can easily 
understand the difference with the results of Ref.\cite{sils}, quoted 
in Table \ref{tabIV}, due to the larger strong coupling constant used in these works. 
This is due to an obvious renormalization of the gluon exchange parameters in order to reproduce 
the meson spectra in two different formalisms (gluon exchange alone and gluon
exchange + boson exchanges).

There is an important aspect that should be emphasized at this point. For the
predicted bound systems the thresholds are determined by D and B mesons. Such
mesons are described just in terms of a confining and a one-gluon exchange 
interaction [see Eq. (\ref{pot})]. Therefore, one could use the same restricted
interaction to calculate the total energy of the tetraquark, but one
would obtain different results (see Table \ref{tabIV}). The reason for this
discrepancy stems from the fact that in this case there also appear interactions
between light quarks that need to be described by means of a more elaborated
interaction. As we have previously mentioned this shows the importance of reproducing the full meson spectrum
(and not a reduced set of states) to make thoughtful predictions.

The structure of our interaction also allows for a study of the influence of the
different color configurations in the tetraquarks binding energy. In Table
\ref{tabV} we present the results for the probabilities of the two color
components $\{\bar33\}$ and $\{6\bar6\}$ in the tetraquark wave function. We
observe that for the bound states channel the probability of $\{6\bar6\}$ is
almost negligible, but its influence increases when the heavy quark mass
decreases. Similar results have also been reported in the context of a pure
one-gluon exchange model \cite{mar}. However, for unbound channels this
probability tends to increase reaching values of 25\% for the $qq\bar s \bar s$
$(S,I)=(0,1)$. This would be an indication that the $\{6\bar6\}$ channel may be 
important for the pure light
tetraquark sector and therefore should not be neglected a priori.

The dominance of the $\{\bar3 3\}$ channel has a more clear physical interpretation 
if we change the color basis from
$\{\mid \overline{3} _{12},3_{34}>,\mid 6_{12},\overline{6}_{34}>\}$ 
to $\{\mid 1_{13},1_{24}>,\mid 8_{13},8_{24}>\}$. 
As stated above, the second basis is more
appropriate to describe asymptotic channels 
because the first vector describes
two physical mesons, whereas the second one 
describes two colored meson states.
The relation between them is given by 

\begin{eqnarray}
&\mid &\overline{3}_{12},3_{34}>=\sqrt{\frac{1}{3}}\mid
1_{13},1_{24}>-\sqrt{\frac{2}{3}}\mid 8_{13},8_{24}> 
\nonumber \\
&\mid &6_{12},\overline{6}_{34}>=\sqrt{\frac{2}{3}}\mid
1_{13},1_{24}>+\sqrt{\frac{1}{3}}\mid 8_{13},8_{24}>
\end{eqnarray}

\noindent Being the $\mid \overline{3}_{12},3_{34}>$ channel 
the dominant one, the contribution to the tetraquark 
binding energy of the colored meson channel is
more important than the two physical meson states. 
This fact can be interpreted in
the sense that our tetraquarks are a more 
complicated object than a pure meson-meson resonance.

Let us finally discuss the possible decay channels of these structures.
The allowed decay modes depend on the relationship between the tetraquark mass
and the sum of the masses of the possible decay products. In the case of the
charmed tetraquark the results shown in Table \ref{tabIII} satisfy
$m_D+m_{D^*}\geq m_{(qq)(\bar c \bar c)}\geq 2m_D$, and therefore the strong
decay is forbidden, being the most probable decay the electromagnetic
process $T[(qq)(\bar c \bar c)]\rightarrow D^+ D^0 \gamma$, that could be identified 
detecting a relatively soft photon, $E_\gamma\leq 140$ MeV. In the case of 
the $qq\bar b \bar b$ structure we obtain (see Table \ref{tabIII}) 
$m_{(qq)(\bar b \bar b)}-2m_B< 0$, 
and therefore it can only decay through two consecutive weak processes.

\section{Summary}

We were faced in this work with the
possible existence of tetraquark structures with heavy-light
flavors in a variational approach. 
To be as much predictive as possible we
have started from a chiral constituent quark model 
which is able to describe the $NN$ interaction and also gives a reasonable
description of the non-strange
baryon spectrum. The generalization to strange and heavy flavors has 
been asked to produce
a nice fit of the meson spectrum. We have proved that 
for the tetraquark description it is enough
to consider a simplified variational wave function 
depending only on the square of the Jacobi
coordinates except for those cases for which
this wave function is forbidden by the Pauli
principle. We have found that the influence
of the $\{6\overline{6}\}$ color
channel could be important in the light sector. For the heavy flavors its contribution
is almost negligible which suggests a non-trivial color structure
for the tetraquark in terms of meson-meson components.

Our calculation predicts the existence of only one bound tetraquark
with $(S,I)=(1,0)$ in both the bottom and the charm sectors. While the
first one should decay weakly, the second one would do it electromagnetically.

\section{acknowledgments}
This work has been partially funded by
Ministerio de Ciencia y Tecnolog{\'\i}a
under Contract No. BFM2001-3563, by Junta de
Castilla y Le\'{o}n under Contract No. SA-109/01,
and by a IN2P3-CICYT agreement.

\begin{table}[tbp]
\caption{Quark-model parameters.}
\label{tabI}
\begin{tabular}{cccc|ccc|ccc}
& $a_c$\,(MeV) & 430.0 &  & $m_S$\,(fm$^{-1}$) & 3.42 &  & $ m_{u,d}$\,(MeV) & 313 &  \\ 
& $\mu_c$\,(fm$^{-1}$) & 0.7 &  & $m_{PS}$\,(fm$^{-1}$) & 0.7 &  & $m_s$ \,(MeV) & 555 &  \\ 
& ${g^2_{ch}/{4\pi}}$ & 0.54&  & $m_K$\,(fm$^{-1}$) & 2.509 &  & $m_c$\,(MeV) & 1752 &  \\ 
& $\alpha_{\,0}$ & 2.1181&  & $m_{\eta}$\,(fm$^{-1}$) & 2.772 &  & $ m_b $\,(MeV) & 5100 &  \\ 
& $\Lambda_0$\,(fm$^{-1}$) &0.113178 &  & $\Lambda_{PS,S}$\,(fm$^{-1}$) & 4.2 & & $\theta_p$\,(rad) &$-$0.2618 &  \\
& $\mu_0$ (MeV)&36.9762&  &$\Lambda_{K,\eta}$\,(fm$^{-1}$)&5.2& &$\hat r_0\,$(fm\,MeV)&28.17&\\
& & & &$\Delta$\,(MeV)&181.1&&&&
\end{tabular}
\end{table}

\begin{table}[tbp]
\caption{Meson spectra for the model described on the text.  
Experimental data are taken from Ref. \protect\cite{dat}.
All masses are given in MeV.}
\label{tabII}
\begin{tabular}{cccccccccc}
& Meson & $\pi$ & $\eta$ & $\rho$ & $\eta^{\prime}$ & $\phi$ & $K$
& $K^*$ \\ 
& Result & 139.5 & 571.6 & 771.6 & 955.7 & 1020.4 & 496.0 & 910.5 &  \\ 
& Experiment & 138.04 & 547.30 & 771.10 & 957.78 & 1019.46 & 
495.0 & 891.66  \\ \hline
& Meson & $D$ & $D^*$ & $\eta_c$ & $J/\psi$ & $B$ & $B^*$ & $\Upsilon$ \\ 
& Result & 1883.2 & 2008.9 & 2989.9 & 3097.0 & 5280.8 & 5321.2 & 9505.0 
\\ 
& Experiment & 1867.7 & 2008.9 & 2979.7 & 3096.87 & 5279.2 & 
5325.0 & 9460.30
\end{tabular}
\end{table}

\begin{table}[tbp]
\caption{$E_T(qq \bar Q \bar Q)$ and $\Delta E$ energies in MeV.}
\label{tabIII}
\begin{tabular}{cccccccccc}
& $(S,I)$  &  & (0,0) & (0,1) & (1,0) & (1,1) & (2,0) & (2,1) &  \\ \hline
& $[(qq)(\bar s\bar s)]$ & $E_T$ & 2239 & 1804 & 1528 & 1940 & 2681 & 2020 & \\ 
&  & $\Delta E$ & +1247 & +812 & +126 & +538 & +860 & +199 &  \\ 
& $[(qq)(\bar c\bar c)]$ & $E_T$ & 4351 & 4150 & 3764 & 4186 & 4849 & 4211 & \\ 
&  & $\Delta E$ & +585 & +384 & $-$129 & +293 & +830 & +192 &  \\ 
& $[(qq)(\bar b\bar b)]$ & $E_T$ & 10820 & 10690 & 10261 & 10698 & 11350 & 10707 &  \\ 
&  & $\Delta E$ & +258 & +128 & $-$341 & +96 & +708 & +65 & 
\end{tabular}
\end{table}

\begin{table}[tbp]
\caption{$E_T (q q \bar{Q} \bar{Q})$ 
and $\Delta E$ in MeV neglecting the pseudoscalar part of the Goldstone boson exchanges.} 
\label{tabIV}
\begin{tabular}{cccccccccc}
& $(S,I)$  &  & (0,0) & (0,1) & (1,0) & (1,1) & (2,0) & (2,1) &  \\ \hline
& $[(qq)(\bar s\bar s)]$ & $E_T$ & 2630 & 1999 & 1996 & 2139 & 2772 & 2186 & \\ 
&& $\Delta E$ & +1638 & +1007 & +594 & +737 & +951 & +365 &  \\ 
& $[(qq)(\bar c\bar c)]$ & $E_T$ & 4666 & 4175 & 4101 & 4231 & 4851 & 4254 & \\ 
&& $\Delta E$ & +900 & +409 & +208 & +338 & +832 & +235 &  \\ 
&&$\Delta E$ from Ref.\cite{sils} & & & +19 & & &  \\ 
& $[(qq)(\bar b\bar b)]$ & $E_T$ & 11136 & 10735 & 10612 & 10743 & 11352 & 10752 &  \\ 
&& $\Delta E$ & +574 & +173 & +10 & +141 & +710 & +110 & \\
&&$\Delta E$ from Ref.\cite{sils} & & & $-$131 &+56 &&+30 &  
\end{tabular}
\end{table}

\begin{table}[tbp]
\caption{Probability of the different color channels.}
\label{tabV}
\begin{tabular}{ccccccc}
& $(S,I)$   & \multicolumn{2}{c}{(0,1)} & \multicolumn{2}{c}{(1,0)}   \\ 
\hline
&& $P_{\{\bar3 3\}}$ & $P_{\{6\bar6\}}$ & $P_{\{\bar3 3\}}$ & $P_{\{6\bar6\}}$  \\ 
\hline 
& $[(qq)(\bar s\bar s)]$&0.7469&0.2531&0.9706&0.0294 \\ 
& $[(qq)(\bar c\bar c)]$&0.8351&0.1649&0.9940&0.060  \\ 
& $[(qq)(\bar b\bar b)]$ &0.9895&0.0105&0.9993&0.0007 \\
\end{tabular}
\end{table}

\bigskip


\begin{references}

\bibitem{gell} M. Gell-Mann, 
Phys. Lett. {\bf 8}, 214 (1964).

\bibitem{jaf} R.L. Jaffe, 
Phys. Rev. D {\bf 15}, 267 (1977).

\bibitem{new} M.A. Moinester, 
Z. Phys. A {\bf 355}, 349 (1996).

\bibitem{mar} 
S. Zouzou, B. Silvestre-Brac, C. Gignoux and J.M. Richard, Z. Phys. C {\bf 30}, 457 (1986).
L. Heller and J.A. Tjon, Phys. Rev. D {\bf 35}, 969 (1987). 

\bibitem{boris} B.A. Gelman and S. Nussinov,  Phys. Lett. B {\bf 551}, 296
(2003).

\bibitem{selex} M. Mattson {\it et al.} (SELEX collaboration), Phys. Rev. Lett.
{\bf 89}, 112001 (2002).

\bibitem{hell} J. Carlson, L. Heller, and J.A. Tjon, 
Phys. Rev. D {\bf 37}, 744 (1988).

\bibitem{wise} A.V. Manohar and M.B. Wise, 
Nucl. Phys. B {\bf 399}, 17 (1993).

\bibitem{sils} B. Silvestre-Brac and C. Semay, 
Z. Phys. C {\bf 57}, 273 (1993); {\it ibid} {\bf 59}, 457 (1993); {\it ibid} {\bf 61}, 271 (1994).

\bibitem{bhad} R.K. Bhaduri, L.E. Cohler, and Y. Nogami,
Nuov. Cim. A {\bf 65}, 376 (1981).

\bibitem{rich} S. Pepin, Fl. Stancu, M. Genovese, and J.-M. Richard, 
Phys. Lett. B {\bf 393}, 119 (1997).

\bibitem{paco} F. Fern\'andez, A. Valcarce, U. Straub, and A. Faessler,
J. Phys. G {\bf 19}, 2013 (1993). D.R. Entem, F. Fern\'{a}ndez, 
and A. Valcarce, Phys. Rev. C {\bf 62}, 034002 (2000).

\bibitem{bruno} B. Juli\'a-D\'{\i}az, J. Haidenbauer, A. Valcarce,
and F. Fern\'andez, Phys. Rev. C {\bf 65}, 034001 (2002). 
H. Garcilazo, A. Valcarce, and F. Fern\'andez, Phys. Rev. C {\bf 64},
058201 (2001).

\bibitem{luis} L.A. Blanco, F.Fern\'andez, and A. Valcarce,
Phys. Rev. C {\bf 59}, 428 (1999).

\bibitem{diak} D.I. Diakonov and V.Yu. Petrov, 
Nucl. Phys. B {\bf 245}, 259 (1984); {\it ibid} {\bf 272}, 457 (1986).

\bibitem{scad} M.D. Scadron, Phys. Rev. D {\bf 26}, 239 (1982).

\bibitem{DeR} A. de R\'ujula, H. Georgi, and S.L. Glashow, 
Phys. Rev. D {\bf 12}, 147 (1975).

\bibitem{isg2} S. Godfrey and N. Isgur, Phys. Rev. D {\bf 32}, 189 (1985).

\bibitem{sim} Yu. A. Simonov, Yad. Fiz. {\bf 58}, 113 (1995).

\bibitem{pre2} T.A. L\"ahde, C.J. Nyf\"alt, and D.O. Riska, Nucl. Phys. A {\bf
645}, 587 (1999).

\bibitem{coup} A.M. Badalian and D.S. Kuzmenko, Phys. Rev. D {\bf 65}, 016004 (2001).

\bibitem{pre1} C.T.H. Davies, K. Hornbostel, G.P. Lepage, P. McCallum, J.
Shigemitsu, and J. Sloan, Phys. Rev. D {\bf 56}, 2755 (1997).

\bibitem{scre} G.S. Bali, Phys. Rep. {\bf 343}, 1 (2001).

\bibitem{bumu} J. Burger, R. M\"uller, K. Tragl, and H.M. Hofmann,
Nucl. Phys. A {\bf 493}, 427 (1989).

\bibitem{minu} F. James and M. Roos, 
Comput. Phys. Commun. {\bf 10}, 343 (1973).

\bibitem{isg} J. Weinstein and N. Isgur, 
Phys. Rev. D {\bf 27}, 588 (1983).

\bibitem{dat} K. Hagiwara {\it et al.}, 
Phys. Rev. D {\bf 66}, 010001 (2002).

\bibitem{brink} D.M. Brink and Fl. Stancu, 
Phys. Rev. D {\bf 57}, 6778 (1998).

\end{references}
\end{document}